\documentstyle[psfig,conf_iap]{article}

\def\etal{et\ al.}
\def\CIV{C~{\sc iv}}
\def\MgII{Mg~{\sc ii}}
\def\HI{H~{\sc i}}
\def\hMpc{$h^{-1}$~Mpc}

\begin{document}

\heading{The Clustering of \CIV\ and \MgII\ Absorption--Line Systems}
 
\par\medskip\noindent

\author{Jean M. Quashnock$^{1}$, Daniel E. Vanden Berk$^{1,2}$}

\address{Dept. of Astronomy \& Astrophysics, University of Chicago, 
Chicago, IL 60637}

\address{Dept. of Astronomy, University of Texas at Austin, 
Austin, TX 78712}

\begin{abstract}
We have analyzed the clustering of \CIV\ and \MgII\
absorption--line systems on comoving scales from 1 to 16 \hMpc ,
using an extensive catalog of heavy--element QSO absorbers
with mean redshift $\langle z\rangle_{\rm \CIV} = 2.2$ 
and $\langle z\rangle_{\rm \MgII} = 0.9$.
For the \CIV\ sample as a whole,
the absorber line--of--sight correlation function 
is well fit by a power law of the form
$\xi_{\rm aa}(r)={\left(r_0/r\right)}^\gamma$,
with maximum--likelihood values of $\gamma = 1.75\,^{+0.50}_{-0.70}$
and comoving $r_0 = 3.4\,^{+0.7}_{-1.0}$ \hMpc\ ($q_0=0.5$).
This clustering is of the {\em same form} as that for galaxies and clusters
at low redshift, and of amplitude such that absorbers are
correlated on scales of clusters of galaxies.
We also trace the {\em evolution} of the mean amplitude 
$\xi_0(z)$ of the correlation function from $z=3$ to $z=0.9$.
We find that, when parametrized in the conventional manner as 
$\xi_0(z)\propto (1+z)^{-(3+\epsilon)+\gamma}$,
the amplitude grows {\em rapidly}
with decreasing redshift, with maximum--likelihood value for the
evolutionary parameter of $\epsilon = 2.05 \pm 1.0 $ ($q_0=0.5$).
The rapid growth seen in the clustering of absorbers
is consistent with gravitationally induced growth of perturbations.
\end{abstract}

\section{Introduction}

In a previous paper \cite{Quash96}, Quashnock, Vanden Berk, \& York
analyzed line--of--sight correlations of 
\CIV\ and \MgII\ absorption--line systems on large scales,
using an extensive catalog \cite{Y97}\ of 2200 
heavy--element absorption--line systems in over 500 QSO spectra.
Here, we extend that analysis
to smaller comoving scales --- from 1 to 16 \hMpc\ ---
and relate the small--scale clustering of absorbers 
to galaxy clustering in general.

The \CIV\ and \MgII\ data sample is drawn from the catalog of 
Vanden Berk \etal\ \cite{Y97},
using the same selection criteria as those in \cite{Quash96}.
It consists of 260 \CIV\ absorbers, drawn from 202 lines of sight,
with redshifts ranging from $1.2 < z < 3.6$ and
mean redshift $\langle z\rangle_{\rm \CIV} = 2.2$,
and 64 \MgII\ absorbers, drawn from 278 lines of sight,
with redshifts ranging from $0.3 < z < 1.6$ and
mean redshift $\langle z\rangle_{\rm \MgII} = 0.9$.

Unless otherwise noted, we take $q_0=0.5$ and $\Lambda=0$.
We follow the usual convention and take the Hubble constant
to be 100\,$h$\ km~s$^{-1}$~Mpc$^{-1}$.
A more detailed version of this work, 
including more results and outlining our maximum--likelihood method,
has appeared elsewhere \cite{Quash97}.

\begin{figure}
\centerline{\vbox{
\psfig{figure=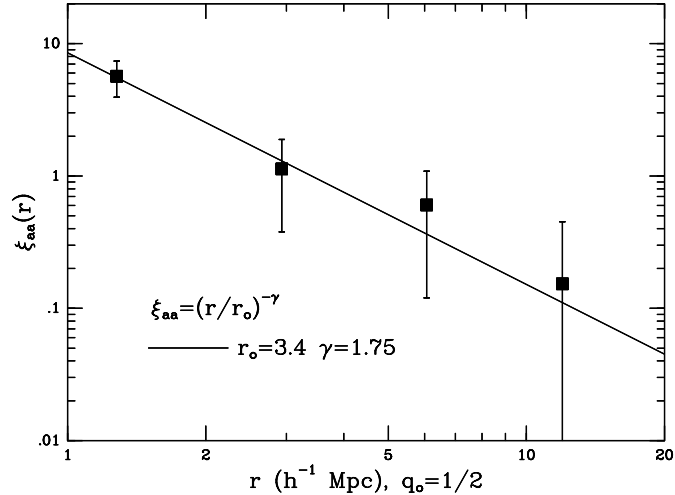,height=6.5cm,angle=-90}
}}
\caption[]{Line--of--sight correlation function, 
$\xi_{\rm aa}(r)$, for the entire sample of \CIV\ absorbers, 
as a function of absorber comoving separation, $r$, 
in 4 logarithmic bins from 1 to 16 \hMpc .
The vertical error bars through the data points 
are 1-$\sigma$ errors in the  estimator for $\xi_{\rm aa}$.
Also shown is a power--law fit
of the form $\xi_{\rm aa}(r)={\left(r_0/r\right)}^\gamma$,
with maximum--likelihood values $\gamma = 1.75$
and comoving $r_0 = 3.4$ \hMpc\ ($q_0=0.5$).}
\end{figure}

\begin{figure}
\centerline{\vbox{
\psfig{figure=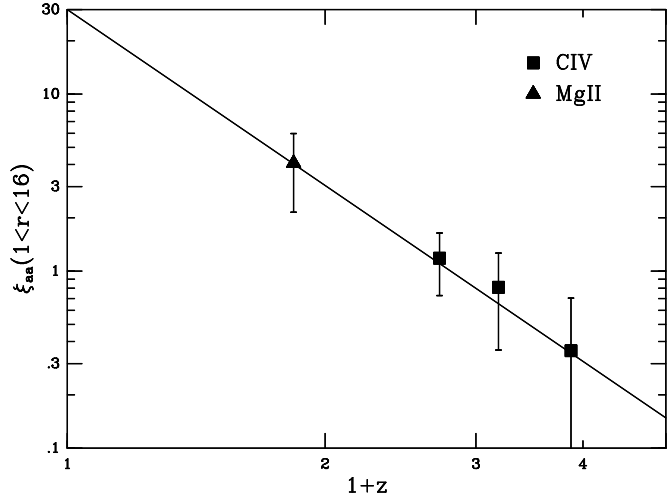,height=6.5cm,angle=-90}
}}
\caption[]{Mean correlation function,
$\xi_0(z)$, averaged over comoving scales 
$r$ from 1 to 16 \hMpc, as a function of redshift.
Shown are values for the low ($1.2<z<2.0$), medium ($2.0<z<2.8$), 
and high ($2.8<z<3.6$) redshift \CIV\ sub--samples,
as well as for the \MgII\ sample ($0.3<z<1.6$).
The solid line is a maximum--likelihood fit of the form
$\xi_0(z)\propto (1+z)^{-(3+\epsilon)+\gamma}$,
with $\epsilon = 2.05$ and $\gamma=1.75$ ($q_0=0.5$).}
\end{figure}

\section{Form and Evolution of the Correlation Function}

Figure~1 shows the 
line--of--sight correlation function $\xi_{\rm aa}(r)$,
for the entire sample of \CIV\ absorbers
(with mean redshift $\langle z\rangle_{\rm \CIV} = 2.2$),
as a function of absorber comoving separation $r$ from 1 to 16 \hMpc ,
in 4 octaves.
The vertical error bars through the data points 
are 1-$\sigma$ errors in the  estimator for $\xi_{\rm aa}$.
The correlation function and error bars are
computed in the same fashion and using the same selection criteria
as those in \cite{Quash96},
except that we have combined all absorbers lying within 1.0 (instead of 3.5)
comoving \hMpc\ of each other into a single system.

Using the maximum--likelihood method of \cite{Quash97},
we find that,  for the \CIV\ sample as a whole,
the line--of--sight correlation function
is well described by a power law of the form
$\xi_{\rm aa}(r)={\left(r_0/r\right)}^\gamma$, with maximum--likelihood
values of $\gamma = 1.75\,^{+0.50}_{-0.70}$
and comoving correlation length $r_0 = 3.4\,^{+0.7}_{-1.0}$ \hMpc\ ($q_0=0.5$).
The clustering of absorbers at high redshift is thus 
of the {\em same form} as that found for galaxies and clusters
at low redshift ($\gamma = 1.77 \pm 0.04$ for
galaxies \cite{DP83}, $\gamma = 2.1 \pm 0.3$ for clusters \cite{Nichol92}),
and of amplitude such that absorbers are
correlated on scales of clusters of galaxies.
It appears that the absorbers are tracing the large--scale structure
seen in the distribution of galaxies and clusters, and are doing so 
at high redshift. The finding strengthens the case for using absorbers
in probing large--scale structure.

We have investigated the evolution of the clustering of absorbers
by dividing the \CIV\ absorber sample into three 
approximately equal redshift sub--samples,
and comparing these to the \MgII\ sample.
Figure~2 shows the mean of the correlation function,
$\xi_0(z)$, averaged over comoving scales 
$r$ from 1 to 16 \hMpc,
for the low ($1.2<z<2.0$), medium ($2.0<z<2.8$), 
and high ($2.8<z<3.6$) redshift \CIV\ sub--samples,
as well as for the \MgII\ sample ($0.3<z<1.6$).
The amplitude of the correlation function is clearly growing
{\em rapidly} with decreasing redshift.

We have used the maximum--likelihood formalism of \cite{Quash97}\
to describe the evolution of the correlation function.
We have fixed $\gamma$ at its maximum--likelihood value of 1.75
in our analysis, and parametrized the amplitude
of the correlation function in the usual manner as
$\xi_0(z) \propto (1+z)^{-(3+\epsilon)+\gamma}$,
where $\epsilon$ is the evolutionary parameter \cite{Efstat91,GP77}.
Using all the data sets,
we find that the rapid growth is reflected in a large value for
the evolutionary parameter, namely $\epsilon = 2.05 \pm 1.0$.
This value is 3.3-$\sigma$ from the no--evolution value
($\epsilon = -1.25$); thus, at the 99.95 \% confidence level,
growth of the correlation function has been detected.
(These results are with $q_0=0.5$.
With $q_0=0.1$, we expect our estimate of $\epsilon$
to decrease by about 1.3 .)

The rapid growth in the correlation function,
and the correspondingly large value of the evolutionary parameter 
($\epsilon = 2.05 \pm 1.0$) that is implied,
is what is expected in a critical universe ($\Omega_0 = 1$),
both from linear theory of gravitational instability
\cite{Peeb80,Peeb93}, with $\xi \propto {\left(1+z\right)}^{-2}$
(or $\epsilon = 0.75$, if $\gamma=1.75$),
and from numerical simulations \cite{Carlberg97,Colin97}:
For $\Omega_0 = 1$, $\epsilon = 1.0 \pm 0.1$,
whereas for $\Omega_0 = 0.2$, $\epsilon = 0.2 \pm 0.1$.

Evidence for a trend of increasing clustering of Ly$\alpha$
absorbers ($N({\rm\HI}) > 6.3 \times 10^{13}~{\rm cm}^{-2}$)
with decreasing redshift has been found by Cristiani \etal\ \cite{Cr97}.
These authors also find a clear trend of increasing
Ly$\alpha$ absorber clustering with increasing column density,
and find that an extrapolation to column densities
typical of heavy--element systems 
($N({\rm\HI}) >  10^{16}~{\rm cm}^{-2}$)
is consistent with the
clustering observed for \CIV\ absorbers \cite{PB94,SC96}.
Our finding of growth in the clustering
of heavy--element systems with decreasing redshift supports both a
continuity scenario between Ly$\alpha$ and heavy--element systems \cite{Cr97},
and the common action of gravitational instability.

The strong clustering that we find 
in the heavy--element absorption--line systems is
thus not surprising, given that most of the sample
consists of the strongest systems with relatively large
equivalent widths (order 0.4 \AA\ and greater),
and the recent claims \cite{Cr97,Dod97}\ of a strong dependence
of clustering strength on the column density of the systems.
We do confirm that the weaker systems (equivalent widths
0.2 \AA\ and less) are less clustered than the stronger ones, by
a factor of two or so; unfortunately, most of the spectra used
to assemble the Vanden Berk \etal\ catalog \cite{Y97}\ are not of sufficient
quality to yield a large number of weak systems.

\acknowledgements{ }
We wish to acknowledge the long--term direction of Don York,
in compiling the extensive catalog of heavy--element
absorbers used in this study,
and in providing intellectual leadership for the project.

\end{document}